\documentclass[aps,prb,twocolumn,reprint,showpacs,superscriptaddress,floatfix]{revtex4-1}
\usepackage{dcolumn}
\usepackage{bm}
\usepackage[pdftex]{graphicx}

\begin{document}

\title{Frustration effects and role of selective exchange coupling for
magnetic ordering \\
in the Cairo pentagonal lattice}

\author{A. Chainani}
\email{chainani@spring8.or.jp}
\affiliation{RIKEN SPring-8 Centre, Sayo-cho, Sayo-gun, Hyogo 679-5148, Japan}
\author{K. Sheshadri}
\email{kshesh@gmail.com}
\affiliation{226, Bagalur, Bangalore North Taluk, Karnataka - 562149, India}

\date{\today}

\begin{abstract}
The Cairo pentagonal lattice, consisting of an irregular pentagonal tiling of
magnetic ions on two inequivalent sites (3- and 4-co-ordinated ones), 
represents a fascinating example for studying geometric frustration effects in two-dimensions.
In this work, we investigate the spin $S$ = $1/2$ Cairo pentagonal lattice with respect to selective exchange coupling
(which effectively corresponds to a virtual doping of $x$ = $0, 1/6, 1/3$), in a nearest-neighbour antiferromagnetic Ising model. 
We also develop a simple method to quantify geometric frustration in terms of a frustration index $\phi(\beta,T)$, where $\beta$ = $J/\tilde{J}$, the ratio of the two exchange couplings required by the symmetry of the Cairo lattice.
At $T = 0$,  the undoped Cairo pentagonal lattice shows antiferromagnetic ordering  for $\beta \le \beta_{crit}  = 2$, but undergoes a first-order transition to a ferrimagnetic phase for $\beta >$ $\beta_{crit}$. 
The results show that $\phi(\beta,T = 0)$ tracks the transition in the form of a cusp maximum at $\beta_{crit}$. 
 While both phases show frustration, the obtained magnetic structures reveal that the frustration originates in different bonds for the two phases. 
The frustration and ferrimagnetic order get quenched by selective exchange coupling, and lead to robust antiferromagnetic ordering for $x$ = 1/6 and 1/3. From mean-field calculations, we determine the temperature-dependent sub-lattice magnetizations for $x$ = $0, 1/6$ and $1/3$. The calculated results are discussed in relation to known experimental results for trivalent Bi$_2$Fe$_4$O$_9$ and
mixed valent BiFe$_2$O$_{4.63}$. The study identifies the role of frustration effects, the ratio $\beta$ and 
selective exchange coupling for stabilizing ferrimagnetic versus anti-ferromagnetic order in the Cairo pentagonal lattice. 
 
\end{abstract}


\maketitle

\section{Introduction}

Geometric frustration on a lattice is the inability to consistently satisfy all pair-wise spin interactions between lattice sites as defined by a Hamiltonian. The most common example of geometric frustration is the classical nearest-neighbour antiferromagnetic (NNAF) Ising model on a triangular lattice in two dimensions.\cite{Wannier} This seemingly simple case does not show magnetic order down to the lowest temperature, but instead exhibits a finite entropy at $T = 0$.  Interestingly, the more general or complex case, with the Ising-spins replaced by spin 1/2 Heisenberg-spins has been shown to exhibit an ordered ground state, which corresponds to the quantum analogue of the classical Neel ground state.\cite{Bernu} This may be compared with another case of frustration-induced zero-point entropy and absence of ordering, namely, the quantum spin-liquid known for the $S = 1/2$ kagome lattice Heisenberg antiferromagnet.\cite{Depenbrock} These contradictory results i.e. presence or absence of ordering, typify the uncertain role of frustration in two dimensional systems with triangular plaquettes. Extensive studies on a variety of lattice spin models in two and three dimensions have indeed revealed 
and established a plethora of exotic phases and properties due to frustration.\cite{Balents, Mila, Castelnovo}. Well-known examples include resonating valence bond superconductivity\cite{PWA}, 'order by disorder'\cite{Villain}, spin-ice\cite{Ramirez}, spin liquid with charge fractionalization\cite{Fulde}, magnetic monopoles\cite{CastelnovoNature},  and so on. 

Although the triangular plaquette is the smallest unit with intrinsic frustration, any larger plaquette with an odd number of edges (or vertices) would also naturally give rise to frustration. In particular, a pentagonal plaquette also leads to frustration. While a regular pentagonal plaquette cannot form a Bravais lattice, there are 14 known pentagonal tesselations based on irregular pentagons. Early studies\cite{Waldor,Bose,Moessner} discussed frustration effects for the two-dimensional pentagonal lattice obtained from the hexagonal lattice by cutting each hexagon into halves with parallel lines(Fig. 1 of ref. 12). In the following, we call this pentagonal lattice the P1 lattice. Using a transfer matrix approach, Waldor et al. showed that the NNAF Ising model for the P1 lattice had a finite ground state entropy due to frustration.\cite{Waldor} For the NNAF Heisenberg model on the P1 lattice, Bhaumik and Bose identified the possibility of a collinear Neel type ground state order.\cite{Bose} Moessner and Sondhi investigated the P1 lattice in terms of a NNAF Ising model with a transverse field, and they identified a novel sawtooth state consisting of zigzag antiferromagnetic stripes coexisting alternately with frustrated ferromagnetic stripes.\cite{Moessner} Urumov, on the other hand, investigated\cite{Urumov} a different tesellation, the so-called two-dimensional Cairo pentagonal lattice (See Fig. 1). Urumov presented an exact solution of the nearest-neighbour ferromagnetic Ising model for the two dimensional Cairo pentagonal lattice by mapping it onto a Union Jack lattice with nearest and second nearest-neighbour non-crossing interactions.\cite{Urumov} 

However, more recently, since the discovery of an approximate experimental realization of the  Cairo pentagonal lattice in Bi$_2$Fe$_4$O$_9$, which exhibits magneto-electric coupling\cite{Singh} and frustration induced non-collinear magnetism,\cite{Ressouche} there has been a significant resurgence of interest in the  Cairo pentagonal lattice.\cite{Ralko,Rojas, Rousochatzakis,Retuerto,Abakunov,Pchelkina,Nakano}
Ralko discussed the phase diagram of the $XXZ$ spin $S = 1/2$ system under an applied magnetic field and showed that frustration leads to unconventional phases such as a ferrimagnetic superfluid.\cite{Ralko} Rojas et al. employed a direct decoration transformation approach in order to investigate antiferromagnetic as well as ferromagnetic coupling for an exact solution of the Cairo pentaonal lattice.\cite{Rojas} They could thereby show that the phase diagram includes a so called disordered/frustrated state, in addition to ferromagnetic and ferrimagnetic phases. 
In an extensive study of the NNAF Heisenberg model on the Cairo pentagonal lattice, Rousochatzakis et al. revealed the role of an order by disorder mechanism and a possible spin-nematic phase with d-wave symmetry.\cite{Rousochatzakis}
The authors addressed the evolution of the phase diagram from the classical to the quantum limit as a function of $\beta$ = $J/\tilde{J}$, where $J$ and $\tilde{J}$ ( $J_{43}$ and $J_{33}$, respectively, in ref. 20 ) are the two types of exchange couplings which originate in the two types of lattice sites. $J$ is the exchange coupling between a 4-co-ordinated and a 3-co-ordinated site, while $\tilde{J}$ is the exchange coupling between two 3-co-ordinated sites, respectively, of the Cairo pentagonal lattice(see Fig. 1(a) and its caption).

On the experimental front, Retuerto et al. addressed\cite{Retuerto} the role of oxygen non-soichiometry in BiFe$_2$O$_{5-\delta}$ and confirmed that the mixed valence of iron, with a nominal valency of ~Fe$^{3.2+}$ in BiFe$_2$O$_{4.63}$, leads to an antiferromagnetic ground state
with $T_N = 250 K$. In the related system Bi$_4$Fe$_5$O$_{13}$F, Abakunov et al. showed\cite{Abakunov} that the presence of frustrated exchange couplings lead to a sequence of magnetic transitions at $T_1$ = 62 K, $T_2$ = 71 K and $T_N$ = 178 K.
Using neutron diffraction and thermodynamic measurements, the authors could show the formation of a non-collinear antiferromagnet below $T_1$ = 62 K, while the structure between $T_1$ and $T_N$ was partially disordered. Pchelkina and Streltsov carried out ab-initio band structure calculations\cite{Pchelkina} for Bi$_2$Fe$_4$O$_9$
and showed that a complete description required going beyond the two-dimensional Cairo pentagonal lattice. However, considering only the two largest in-plane exchange coupling parameters, the obtained ground state was found to be consistent with experiment. And very recently, Nakano et al. addressed\cite{Nakano} the magnetization process in the two dimensional spin $S = 1/2$ Heisenberg model on the Cairo pentagonal lattice. Using a numerical diagonalization method, they discussed the role of the
ratio of the two types of exchange couplings in driving a quantum phase transition, and very interestingly, found a 1/3 magnetization plateau usually associated with triangular plaquettes. 

The results described above clearly show that frustration effects play an important role
in determining the magnetic ordering in the Cairo pentagonal lattice. However, 
the quantification of frustration
and the role of specific or selective exchange couplings associated with the two types of sites in the lattice has not been addressed yet. Further, while experimental results for the hole-doped system  BiFe$_2$O$_{5-\delta}$ have been reported, it is important to theoretically investigate the role of doping
for magnetic ordering in the Cairo pentagonal lattice. 
In the present study, we address these issues for the spin $S = 1/2$ Cairo pentagonal lattice in a NNAF Ising model.
We first develop a method to quantify geometric frustration in terms of a frustration index $\phi(\beta,T)$.
We find that, for the undoped case (x = 0),  
$\phi(\beta,T=0)$ exhibits a cusp maximum as a function of $\beta$ = $J/\tilde{J}$. Further, in the presence of finite frustration, we identify antiferromagnetic ordering  at low $\beta$, which undergoes a first-order transition to a ferrimagnetic phase for $\beta >$ $\beta_{crit}$ = 2.  
The frustration and ferrimagnetic order get suppressed by selective exchange coupling, and lead to  antiferromagnetic ordering for x = 1/6 and 1/3. Mean-field calculations are carried out to determine the temperature-dependent magnetization for $x$ = 0, 1/6 and 1/3. The results are discussed in relation to known experimental results for trivalent Bi$_2$Fe$_4$O$_9$ and
mixed valent BiFe$_2$O$_{4.63}$. The study identifies the role of frustration effects, the ratio $\beta$ and 
selective exchange coupling in relation to ferrimagnetic versus anti-ferromagnetic ordering in the Cairo pentagonal lattice. 

\begin{figure}
\begin{center}
\includegraphics[width=1\columnwidth]{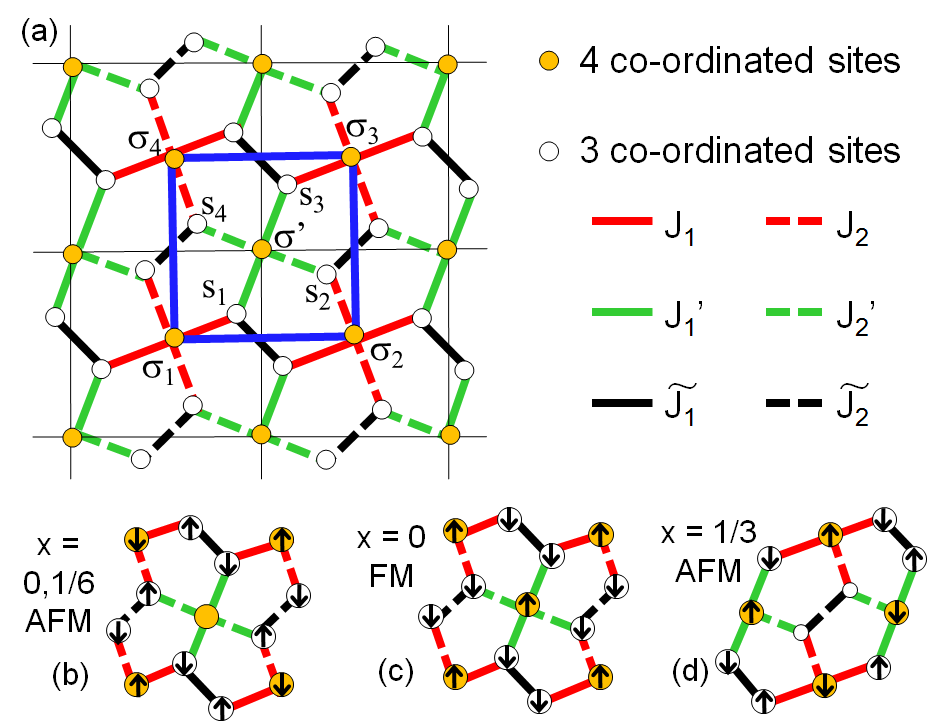}
\end{center}
\caption{(Color online) (a) The unit cell of the two-dimensional Cairo pentagonal lattice(thick blue lines) used in this paper. The undoped ($x$ = 0) case  corresponds to 
$\tilde{J}_1  = \tilde{J}_2$ = $\tilde{J}$, and 
$ J_{1}  = J_1^{\prime} = J_2  = J_2^{\prime} = J$, and the magnetic structure depends on the ratio $\beta$ = $J$/$\tilde{J}$. The  two magnetic structures obtained for $x$ = 0, with finite but different values of frustration index $\phi(T = 0)$, are shown : (b) for $\beta$ = 1, it is antiferromagnetic, and (c) for $\beta$ = 3, it is ferrimagnetic. For $x$ = 1/6, the obtained 'star' antiferromagnetic structure is the same as shown in (b), while (d) shows the 'boat' antiferromagnetic structure obtained for $x$ = 1/3.}
\label{fig1}
\end{figure}

\section{Calculational Details}
\label{method}

The unit cell is as shown in Fig. 1 (thick blue lines) and consists of six Ising spins: four 3-co-ordinated spins $s_1, s_2, s_3, s_4$, a 4-co-ordinated spin $\sigma^{\prime}$ at the center, and another 4-co-ordinated spin $\sigma_1$ at the bottom-left corner. Each of these can take values of $ S = \pm 1/2$. For the most general case, the couplings (that are symmetric in the indices) are denoted as follows:
\begin{eqnarray}
\label{eq:couplings}
\nonumber \\
J_{s_1,s_3} = \tilde{J}_1,&&J_{s_2,s_4} = \tilde{J}_2,  \nonumber \\
J_{s_1,\sigma^{\prime}} = &J_{s_3,\sigma^{\prime}}& = J_1^{\prime},  \nonumber \\
J_{s_2,\sigma^{\prime}} = &J_{s_4,\sigma^{\prime}}& = J_2^{\prime},  \nonumber \\
J_{s_1,\sigma} = &J_{s_3,\sigma}& = J_1,  \nonumber \\
J_{s_2,\sigma} = &J_{s_4,\sigma}& = J_2.
\end{eqnarray}
While the above detailed notations for the couplings are necessary for describing selective exchange couplings, 
which effectively represent virtual doping content ($x$), the undoped case is obtained by setting 
$\tilde{J}_1  = \tilde{J}_2$ = $\tilde{J}$, and 
$ J_{1}  = J_1^{\prime} = J_2  = J_2^{\prime} = J$. As detailed in the following, appropriate limits of these parameters allow us to calculate physical quantities for $x$ = 1/6 and 1/3.
The Ising Hamiltonian is
\begin{equation}
\label{eq:isinglattice}
H = \sum_{i} H_i,
\end{equation}
where
\begin{equation}
\label{eq:isingunitcell}
H_i = \sum_{\alpha} K_{\alpha}^{(i)} s_{\alpha}^{(i)}
\end{equation}
corresponds to the $i^{th}$ unit cell of the two-dimensional Cairo lattice. Here the symbols $K_{\alpha}^{(i)}$ are defined by
\begin{eqnarray}
\label{eq:kvecdefs}
\nonumber \\
K_{1}^{(i)} &=& J_1\sigma^{(i)} + J_1^{\prime}\sigma^{\prime (i)} + \tilde{J}_1 s_3^{(i-y)}, \nonumber \\
K_{2}^{(i)} &=& J_2\sigma^{(i+x)} + J_2^{\prime}\sigma^{\prime (i)} + \tilde{J}_2 s_4^{(i+x)}, \nonumber \\
K_{3}^{(i)} &=& J_1\sigma^{(i+x+y)} + J_1^{\prime}\sigma^{\prime (i)} + \tilde{J}_1 s_1^{(i+y)}, \nonumber \\
K_{4}^{(i)} &=& J_2\sigma^{(i+y)} + J_2^{\prime}\sigma^{\prime (i)} + \tilde{J}_2 s_2^{(i-x)},
\end{eqnarray}
in which the superscripts $i \pm x, ~ i \pm y, ~ i + x + y$ are indices of unit cells neighboring $i$. We perform a mean-field decoupling of the Hamiltonian Eq.(\ref{eq:isinglattice}) according to
\begin{equation}
\label{eq:mft}
S_iS_j \simeq \langle S_i \rangle S_j + \langle S_j \rangle S_i - \langle S_i \rangle \langle S_j \rangle
\end{equation}
for a product of any two spins $S_i$ and $S_j$, where $\langle S \rangle = Tr(Se^{-H_{MF}/k_BT})/Tr(e^{-H_{MF}/k_BT})$ for any $S$. This approximation linearizes the unit cell Hamiltonian Eq.(\ref{eq:isingunitcell}) in the spin variables:
\begin{equation}
\label{eq:mfham}
H_{MF} = c_0 + \sum_{\alpha = 1}^{4} c_{\alpha} s_{\alpha} + c_{5} \sigma^{\prime} + c_6 \sigma.
\end{equation}
 We have suppressed the unit cell index $i$. The coefficients are
\begin{eqnarray}
\label{eq:cdefs}
\nonumber \\
c_0 &=& -2 \tilde{J}_1 m_1 m_3  \cos (k_y) - J_1 m_3 m_6 \cos (k_x + k_y) 
\nonumber \\ && -2 \tilde{J}_2
   m_2 m_4 \cos (k_x) - J_1 m_1 m_6 
\nonumber \\ && - J_2 m_2 m_6 \cos (k_x) - J_2 m_4 m_6 \cos (k_y) 
\nonumber \\ && - (J_1' m_1 +  J_1' m_3 + J_2' m_2 + J_2' m_4) m_5 ,   \nonumber \\
c_1 &=& 2 \tilde{J}_1 m_3  \cos (k_y) + J_1 m_6 + J_1' m_5 ,   \nonumber \\
c_2 &=& (2 \tilde{J}_2 m_4+J_2 m_6) \cos (k_x) + J_2' m_5,   \nonumber \\
c_3 &=& 2 \tilde{J}_1 m_1 \cos (k_y) + J_1 m_6 \cos (k_x + k_y) + J_1' m_5,    \nonumber \\
c_4 &=&  2 \tilde{J}_2 m_2 \cos (k_x) + J_2 m_6 \cos (k_y) + J_2' m_5, \nonumber \\
c_5 &=&  J_1' m_1 + J_1' m_3  + J_2' m_2 + J_2' m_4,  \nonumber \\
c_6 &=& J_1 m_1 + J_2 m_2 \cos (k_x) + J_1 m_3 \cos (k_x + k_y) 
\nonumber \\ && + J_2 m_4 \cos (k_y).  
\end{eqnarray}

\begin{figure}
\begin{center}
\includegraphics[width=1\columnwidth]{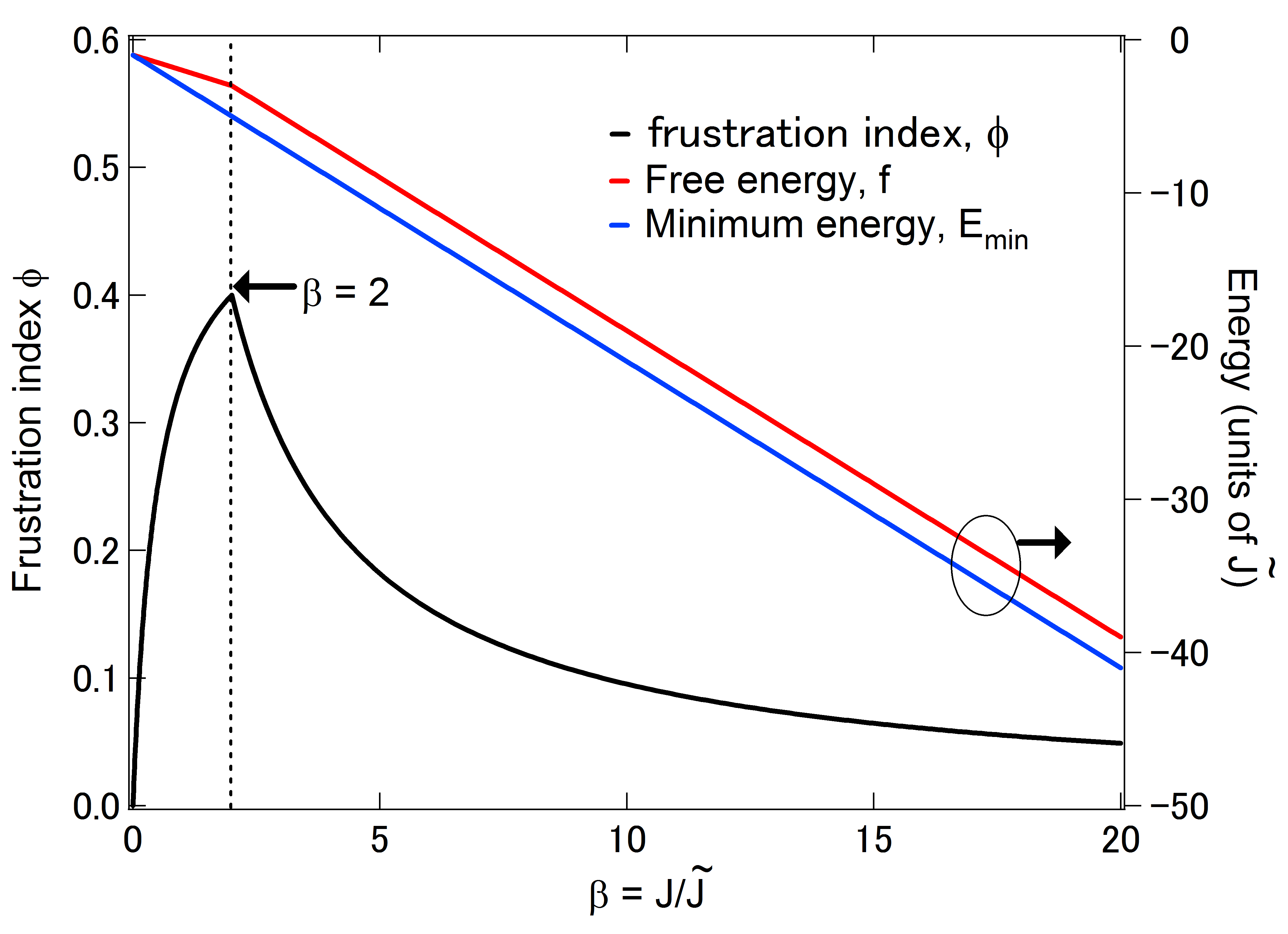}
\end{center}
\caption{(Color online) Plots of (a) frustration index $\phi(\beta,0)$ (Eq. (\ref{eq:phi})), (b) zero-temperature free energy $f(\beta,0)$ (Eq. (\ref{eq:freeenergy})), and (c) $E_{min}(\beta)$ (Eq. (\ref{eq:emin})) as a function of $\beta = J/\tilde{J}$ for the undoped case, $x=0$ ; $\phi(\beta,0)$ exhibits a non-monotonic behaviour and a cusp at $\beta = 2$.}
\label{fig2}
\end{figure}

\begin{figure}
\begin{center}
\includegraphics[width=1\columnwidth]{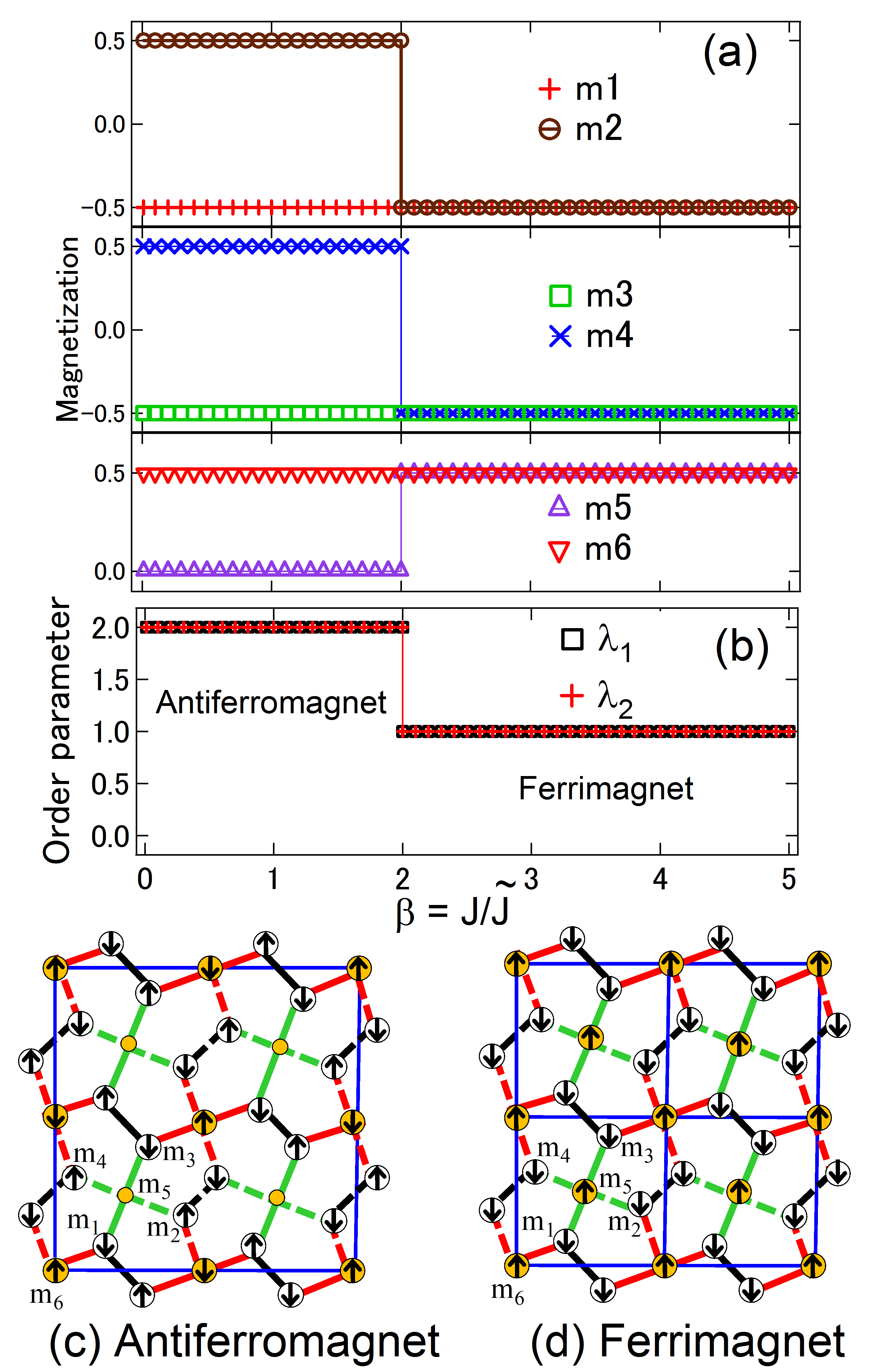}
\end{center}
\caption{(Color online) Components of the order parameter $\mu$ at zero temperature plotted as a function of $\beta$ at $x = 0$. (a) Plots of the sub-lattice magnetizations $m_1$ to $m_6$. The jumps seen for $m_2$, $m_4$ and $m_5$ indicate a change in the magnetic order from an antiferromagnetic phase for $\beta \le 2$ to a ferrimagnetic phase for $\beta>2$.
(b) The SDW lengths $\lambda_1, \lambda_2$ plotted as a function of $\beta$ indicate a doubling of the magnetic lattice along x and y axes for the antiferromagnetic phase for $\beta \le 2$(panel c) compared to the ferrimagnetic phase for $\beta>2$(panel d).}
\label{fig3}
\end{figure}

\begin{figure}
\begin{center}
\includegraphics[width=1\columnwidth]{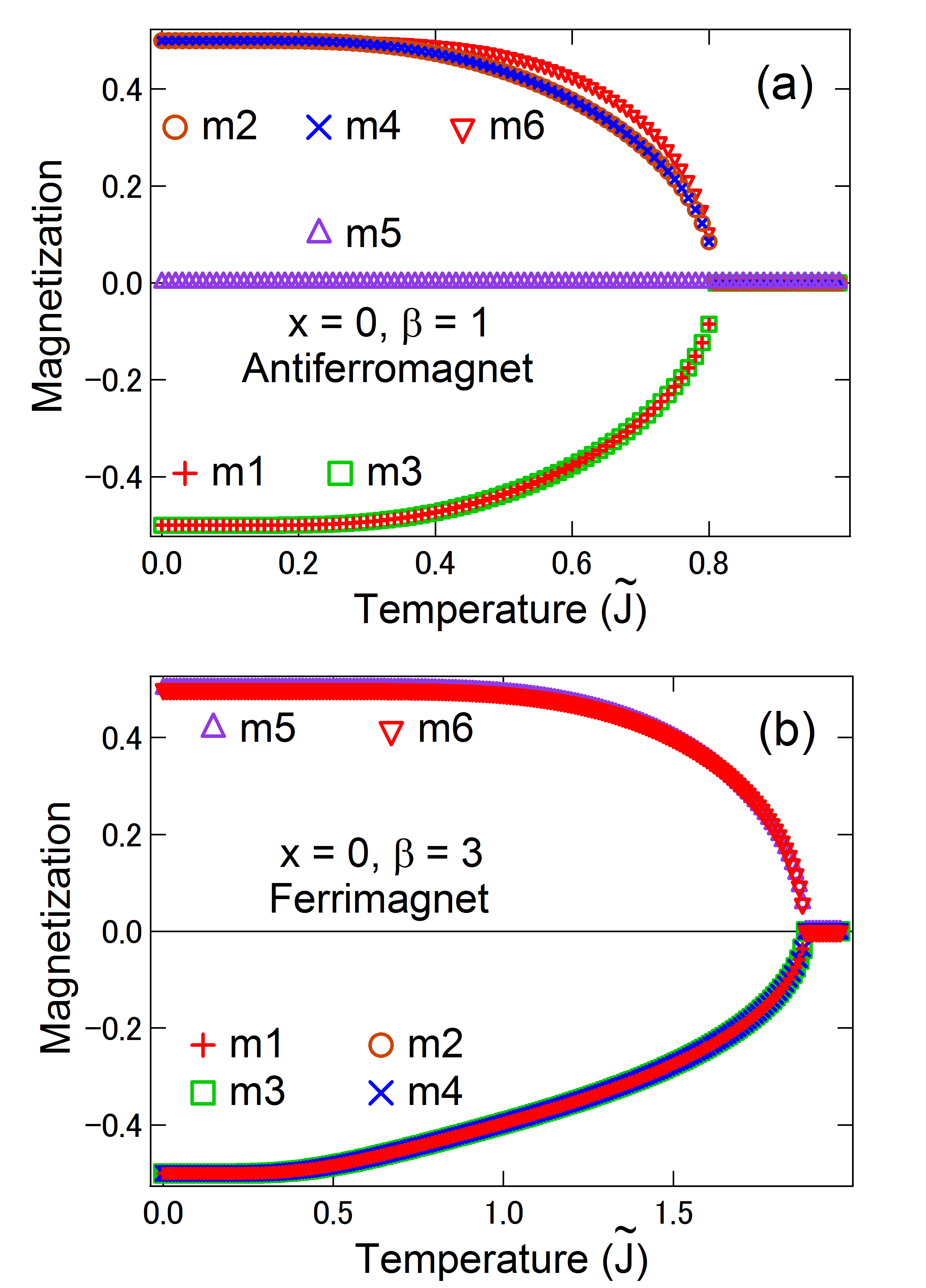}
\end{center}
\caption{Sub-lattice magnetizations $m_i, i = 1-6$, plotted as a function of temperature for the undoped case ($x=0$), for (a) $\beta=1$ (antiferromagnetic phase) and (b) $\beta=3$ (ferrimagnetic phase).}
\label{fig4}
\end{figure}

For the sub-lattice magnetizations, we have introduced the notation $m_{\alpha} = \langle s_{\alpha} \rangle ~ (\mathrm{for} ~ \alpha=1,2,3,4)$, $m_5 = \langle \sigma^{\prime} \rangle, ~ m_6 = \langle \sigma \rangle$. Since the single unit-cell Hamiltonian (Eq.(\ref{eq:isingunitcell})) involves couplings with spins in the neighboring unit cells $i \pm x \pm y$, we have introduced a spin density wave (SDW) vector ${\bf k}$ whose components $k_x=2\pi/\lambda_1, ~ k_y=2\pi/\lambda_2$ appear in the expressions for the coefficients. The six magnetizations $m_{\alpha} ~ (\alpha=1,\cdots,6)$, and the two wave lengths $\lambda_1, \lambda_2$ together form an eight-component mean-field order parameter vector $\mu$:
\begin{equation}
\label{eq:mftop}
\mu = (m_1,~m_2,~m_3,~m_4,~m_5,~m_6,~\lambda_1, ~\lambda_2).
\end{equation}
The thermodynamics of the model is determined by the unit cell free energy $f(T)$:
\begin{equation}
\label{eq:fedef}
e^{-f(T)/k_BT} = Tr (e^{-H_{MF}/k_BT}),
\end{equation}
where the trace is performed over the six spins of the unit cell, each of which takes the values $\pm 1/2$. Therefore
\begin{equation}
\label{eq:freeenergy}
f(T) = c_0 - 6 k_BT\ln(2) - k_BT \sum_{\alpha=1}^{6} \ln \left [\cosh \left ( \frac{c_{\alpha}}{2k_BT} \right )\right ].
\end{equation}
The magnetizations are determined using the self-consistency equations
\begin{equation}
\label{eq:selfcon}
2m_{\alpha} + \tanh\left( \frac{c_{\alpha}}{2k_BT}\right) = 0, ~~~ \alpha=1,\cdots,6,
\end{equation}
while the SDW lengths are determined by free energy minimization $\partial f/\partial\lambda_k = 0, ~ k=1,2$:
\begin{equation}
\label{eq:minfe}
\frac{\partial c_0}{\partial\lambda_k} + \sum_{\alpha=1}^{6} m_{\alpha} \frac{\partial c_{\alpha}}{\partial\lambda_k} = 0, ~~~ k = 1, 2.
\end{equation}
We solve the eight equations in (\ref{eq:selfcon}) and (\ref{eq:minfe}) numerically to obtain the order parameter vector $\mu (\{J_{ij}\}, T)$. 

%
%
We define a frustration index $\phi(\beta,T)$ in the following manner. Consider a reference state for which each bond between neighboring spins were to be fully satisfied. Then, Eq.(\ref{eq:isingunitcell}) would give a zero temperature energy of
\begin{equation}
\label{eq:emin}
E_{min} = - \frac{1}{2} (J_1 + J_1^{\prime} + J_2 + J_2^{\prime} + \tilde{J}_1 + \tilde{J}_2).
\end{equation}
The zero temperature free energy $f(\beta,T=0)$ deviates from $E_{min}(\beta)$ because of frustration, so if we define
\begin{equation}
\label{eq:phi}
\phi(\beta,T) = 1 - \frac{f(\beta,T)}{E_{min}(\beta)},
\end{equation}
then $\phi(\beta,T=0)$ is a convenient measure of frustration. It provides a quantitative measure of frustration as a function of $\beta$ and selective exchange coupling.

\section{Results and Discussions}
\label{results}

\begin{figure}
\begin{center}
\includegraphics[width=1\columnwidth]{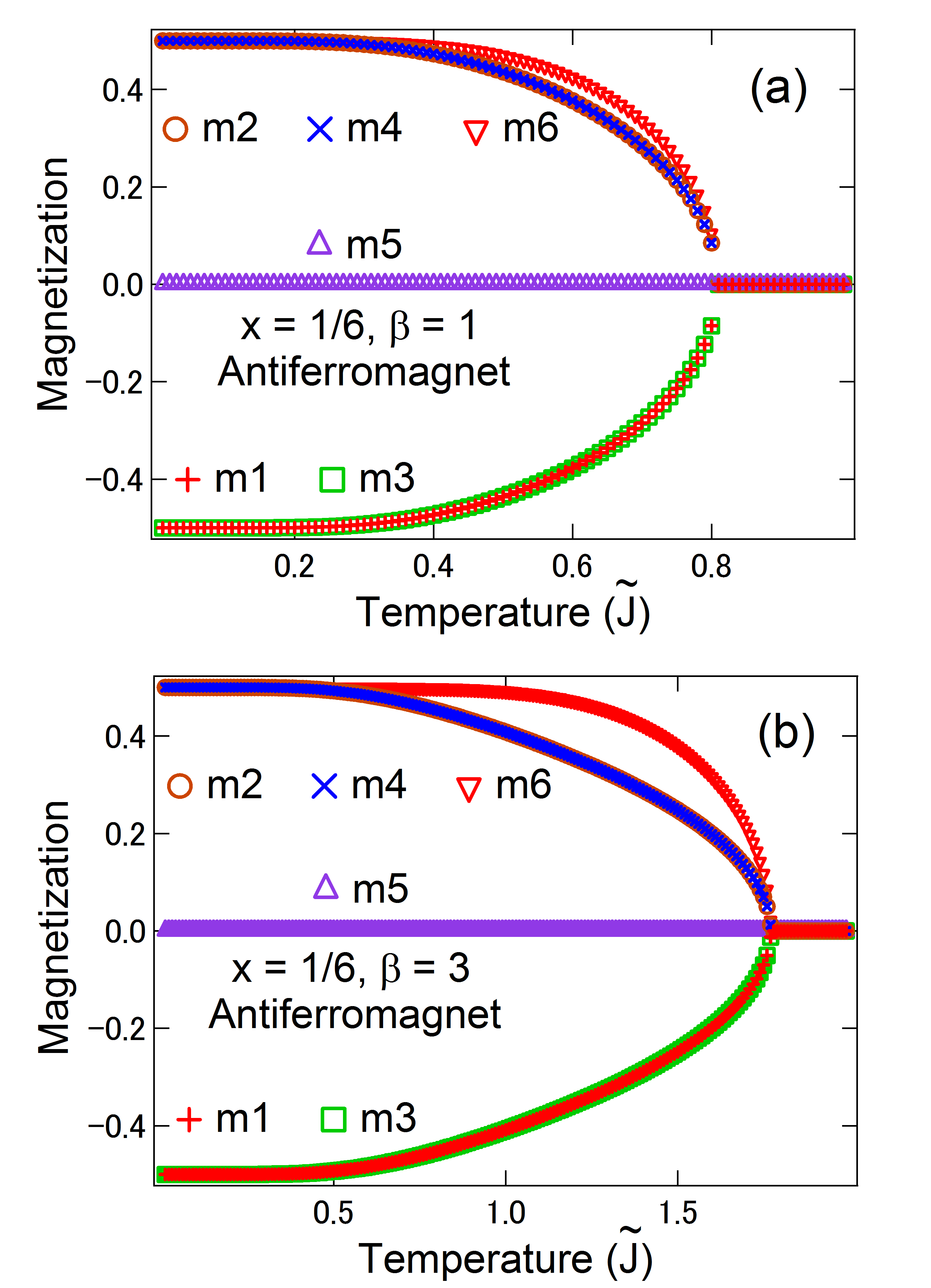}
\end{center}
\caption{Sub-lattice magnetizations $m_i, i = 1-6$, plotted as a function of temperature for the antiferromagnetic phases of $x = 1/6$) for (a) $\beta=1$  and (b) $\beta=3$.}
\label{fig5}
\end{figure}
\begin{figure}
\begin{center}
\includegraphics[width=1\columnwidth]{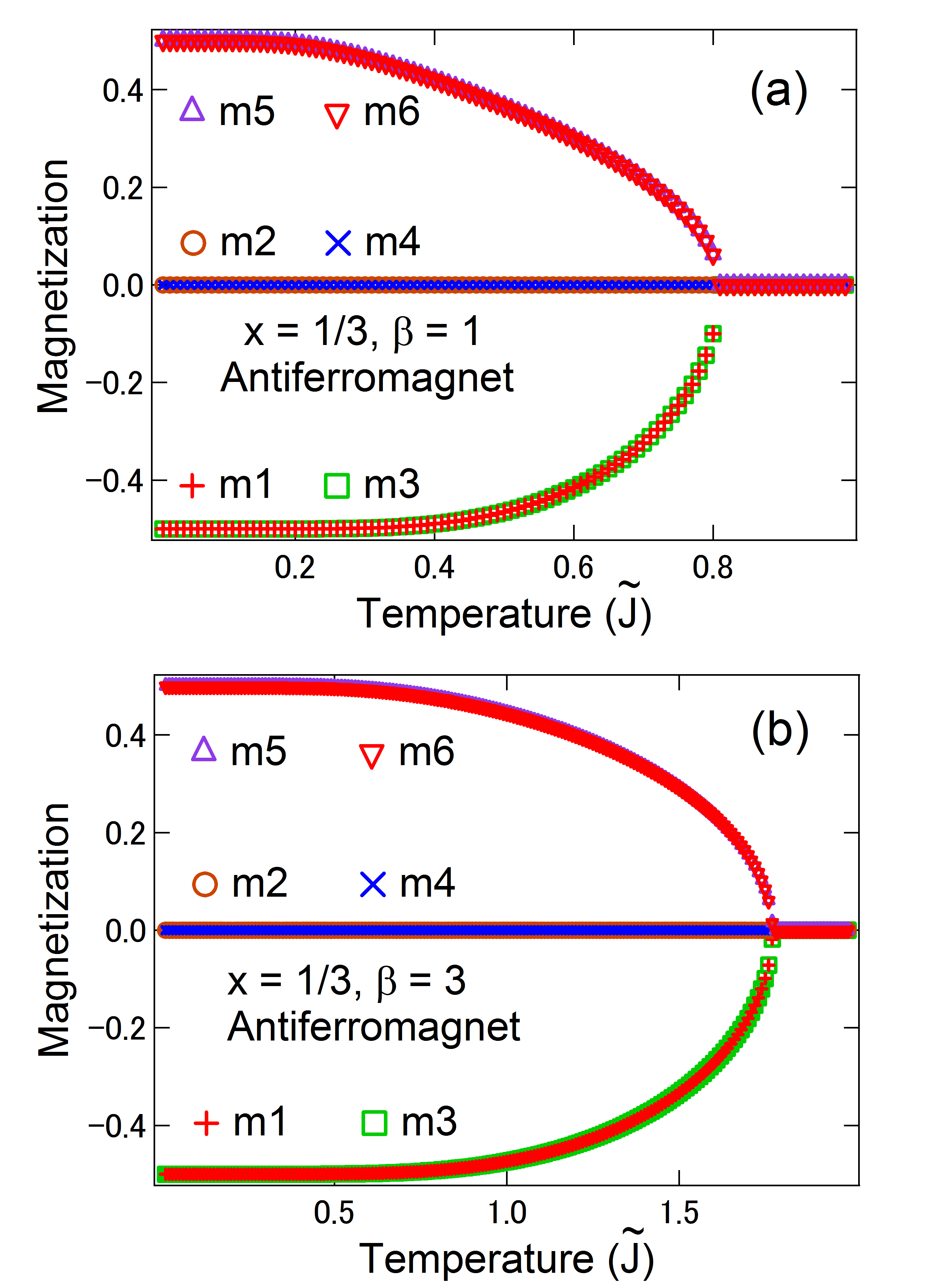}
\end{center}
\caption{Sub-lattice magnetizations $m_i, i = 1-6$, plotted as a function of temperature for the antiferromagnetic phases of $x = 1/3$) for (a) $\beta=1$  and (b) $\beta=3$.}
\label{fig6}
\end{figure}

We first compute the order parameter vector $\mu$ by numerically solving the eight equations in (\ref{eq:selfcon}) and (\ref{eq:minfe}) for the undoped case with $\tilde{J}_1  = \tilde{J}_2$ = $\tilde{J}$, and 
$ J_{1}  = J_1^{\prime} = J_2  = J_2^{\prime} = J$. In Fig. \ref{fig2}, we plot the frustration measure $\phi(\beta,T=0)$ (Eq. (\ref{eq:phi})), the zero-temperature free energy $f(\beta,T=0)$ (Eq. (\ref{eq:freeenergy})), and $E_{min}(\beta)$ (Eq. (\ref{eq:emin})) as a function of $\beta$ = $J/\tilde{J}$. The frustration measure $\phi(\beta,T=0)$ exhibits a clear cusp maximum at $\beta$ = $2$. The cusp originates in the zero-temperature free energy $f(\beta,T=0)$ which exhibits a derivative discontinuity at $\beta$ = $2$. This indicates a first-order transition as a function of $\beta$. As  shown in Fig. \ref{fig3}, the order parameter plots as a function of $\beta$ indicate an antiferromagnetic phase for $\beta \le 2$, which transforms into a ferrimagnetic phase for $\beta>2$. 

We see discontinuities in the sub-lattice magnetizations $m_2$, $m_4$ and $m_5$ at $\beta=2$: while $m_2$ and $m_4$ flip from $1/2$ to $-1/2$, $m_5$ responds by changing from $0$ to $1/2$ to minimize the free energy (Fig. \ref{fig3}(a)). The SDW lengths $\lambda_1, \lambda_2$ also have discontinuities (Fig. \ref{fig3}(b)): $\lambda_1, \lambda_2 = 2$ for $\beta \le 2$, and $\lambda_1, \lambda_2 = 1$ for $\beta>2$. Thus, the magnetic unit cell of the antiferromagnetic phase is doubled along the x and y axes, compared to the ferrimagnetic unit cell as shown in Fig. \ref{fig3}(c) and (d), respectively. 
Interestingly, while the system remains frustrated in both these phases, the frustration index $\phi(T=0) \rightarrow 0$ as $\beta \rightarrow 0$, but goes to a finite value at large $\beta$ asymptotically. Thus, in the limit 
$\beta = 0$, we have isolated dimers on the 3-co-ordinated sites while for $\beta = \infty$, we retain 
the ferrimagnetism.

The ferrimagnetic phase obtained for $\beta>2$ with a total magnetization $M(T = 0) = 1/3$, is identical to earlier studies from (i) exact calculations for the Ising\cite{Urumov,Rojas}, (ii) hard-core bosons\cite{Ralko} and (iii) the Heisenberg\cite{Rousochatzakis} models.
It is noted that, for the antiferromagnetic phase, the sub-lattice magnetization 
$m_5 = \langle \sigma^{\prime} \rangle$ = $0$ for $\beta \le 2$, (see Fig. \ref{fig3}(a)) due to the fact that out of its 4 nearest-neighbour sites, two sites have up-spins and two have down-spins. This leads to an effective cancellation of the magnetization contribution from the $\sigma^{\prime}$ site. In earlier work, Rojas et al.\cite{Rojas} reported a frustrated phase of Ising spins from exact calculations, corresponding to the antiferromagnetic phase obtained from our mean-field calculations, while Rousochatzakis et al.\cite{Rousochatzakis} obtained a so called orthogonal phase for the Heisenberg case.

We now turn to investigating the role of selective exchange coupling at $T=0$. We carry out this exercise mainly to identify the role of specific couplings of the 3- and 4-co-ordinated sites in the Cairo lattice, and to effectively simulate virtual dopings of x = 1/6 and 1/3. From the way we have defined our couplings (see Eq. (\ref{eq:couplings})), it can be observed that if we set $J_1^{\prime} =  J_2^{\prime} = 0$, the central spin $\sigma^{\prime}$ gets disconnected from the lattice. Consequently, the central spin $\sigma^{\prime}$ does not participate in the magnetic ordering and the system effectively represents a virtual hole doping content of $x=1/6$. In the same way, it can be observed that when $J_2 =  \tilde{J}_2 = J_2^{\prime} = 0$,
the spins $s_2$ and $s_4$ get disconnected, corresponding to a virtual hole doping of $x=1/3$. Our results indicate that for both these cases, the system is antiferromagnetic for all $\beta$, i.e. $\lambda_1, \lambda_2 = 2$. 
For $x= 1/6$, the sub-lattice magnetizations are exactly the same as those for $x=0, \beta \le 2$ (Fig. \ref{fig3}), and we therefore refrain from showing these plots in a figure. For $x=1/3$, the actual magnetic structure gets modified as discussed below. Another important difference from the $x=0$ case is that $\phi(\beta,0)=0$ for both values of $x = 1/6$ and $1/3$ and is independent of $\beta$, indicating the absence of frustration. 

The absence of a $\beta$-driven transition for the cases $x=1/6, 1/3$ is easy to understand. In the $x=1/6$ case, the central spin $\sigma^{\prime}$ does not participate in the magnetic ordering. The remaining spins can be looked upon as forming a ``star" configuration with 12 bonds on its periphery, all of which can be satisfied in the antiferomagnetic phase, thus removing frustration fully. An increase of $\beta$ only strengthens the antiferromagnetic order, increasing $T_N$ (see Fig. 7 and associated description). In the $x=1/3$ case, the spins $\sigma_2, \sigma_4$ do not participate in the magnetic ordering. The remaining spins can be looked upon as forming a ``boat" configuration with 10 bonds on its periphery, all of which can be satisfied in the antiferomagnetic phase, again fully removing frustration(Fig. 1(d)). An increase of $\beta$ in this case also, merely strengthens the antiferromagnetic order by increasing $T_N$. While both the $x=1/6$ and $x=1/3$ cases exhibit antiferromagnetic order, the actual spin ordering is found to be different: the former corresponds to a repetition of ``star"(Fig. 1(b)) unit cells, and the latter, a repetition of ``boat"(Fig. 1(d)) unit cells.

\begin{figure}
\begin{center}
\includegraphics[width=1\columnwidth]{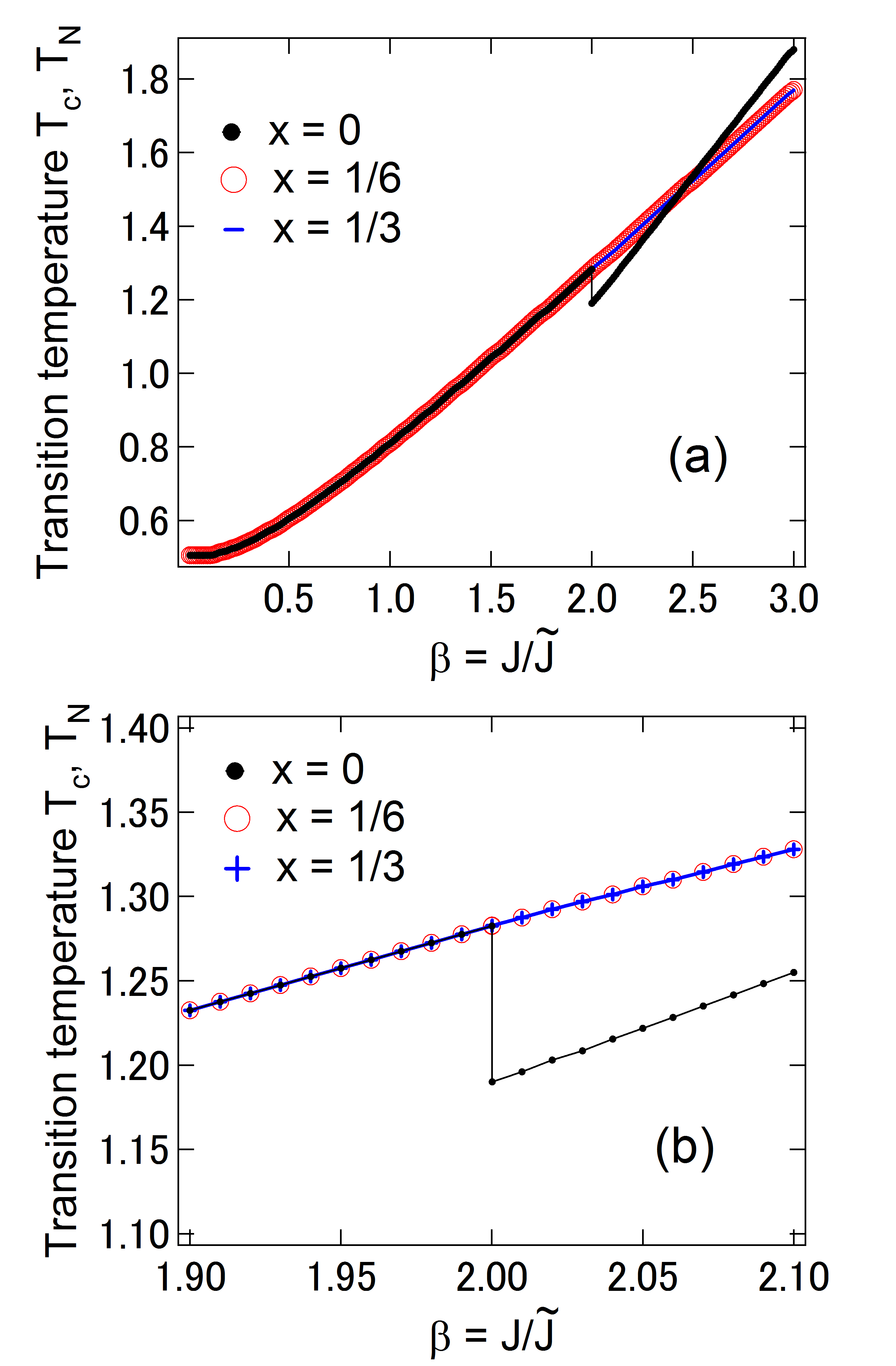}
\end{center}
\caption{(a) Plots of the transition temperature $T_{c}, T_{N}$ as a function of $\beta$ for x = 0, 1/6 and 1/3. A jump in the ordering temperature is seen only for x = 0 at $\beta = 2$. In (b) we present a magnified view of the region near $T_c$.}
\label{fig7}
\end{figure}

In Figs. \ref{fig4}(a) and \ref{fig4}(b), we plot the sub-lattice magnetizations $m_i, i = 1,\cdots,6$ for the undoped case ($x=0$), as a function of temperature in units of $\tilde{J}$ for $\beta=1$ and $\beta=3$, i.e. in the antiferromagnetic and ferrimagnetic phases, respectively. We find that the transition temperature $T_C$ and $T_{N}$(the ferrimagnetic and antiferromagnetic transition temperatures, respectively) for destruction of magnetic order primarily depends on $\beta$. This is further borne out by the temperature-dependence plots for $x=1/6$ presented in Figs. \ref{fig5}(a) and \ref{fig5}(b) for $\beta=1$ and $\beta=3$, respectively. The  temperature-dependence plots for $x=1/3$, presented in Figs. \ref{fig6}(a) and \ref{fig6}(b) for $\beta=1$ and $\beta=3$, respectively, also reflect this. In particular, $T_{N}$ increases on increasing $\beta$ for $x = 1/6$ and $1/3$. In fact, the ordering temperatures increase smoothly with $\beta$ as can be seen
in Fig. \ref{fig7}(a), where we plot $T_C$ and $T_N$  as a function of $\beta$ for the three cases $x=0, 1/6, 1/3$. The discontinuity at the first-order transition for $x=0$ at $\beta=2$ is quite clear and remarkable. In Fig. \ref{fig7}(b), we present a magnified view of the plot around $\beta=2$. 

We can also see in Fig. \ref{fig7}(a) that the dependence of $T_N$ on $x$ is negligible for all $\beta \le 2$. But for $\beta>2$, while the $x=1/6, 1/3$ plots follow the same course as $\beta \le 2$, the $x=0$ curve follows a completely different course because of the discontinuity. The larger $T_N$ of the antiferromagnetic phase compared to the $T_C$ of the ferrimagnetic phase, just above $\beta=2$, suggests a higher stability of the antiferromagnetic phase. However, for $\beta$ values greater than $\sim2.5$, the ferrimagnetic $T_C$ for $x=0$ becomes larger than the antiferromagnetic $T_N$ for $x = 1/6$ and $1/3$. The results also show that $T_N$ goes to a finite value as $\beta \rightarrow 0$, but increases linearly for high $\beta$. 

We now describe the physical picture of the various phases obtained by varying the model parameters. Firstly,
the crucial role that frustration plays in the $\beta$-driven first-order transition is seen from the results of 
the $x = 0$ case. In the $x=0$ antiferromagnet region obtained for $\beta \le 2$, the frustration is confined to the core of the unit cell, i.e. two out of the four $J^{\prime}$-bonds (either the 
 $J_{1}^{\prime}$ or the $J_{2}^{\prime}$ bonds) connecting to the central spin $\sigma^{\prime}$ are frustrated; in this situation, the frustration measure $\phi(0)$ is an increasing function of $\beta$ (see Fig. \ref{fig2}, $\beta \le \beta_{crit}  = 2$ ). With further increase in $\beta$ beyond $\beta_{crit}  = 2$, it becomes energetically favorable to ``eject" the frustration from the core to the $\tilde{J}$ bonds lying on the periphery of the unit cell i.e. the the four $\tilde{J}$-bonds connecting 3-co-ordinated sites become frustrated.
Thus, the $\beta$-driven transition is attended by a change in the location of the frustrated bonds in the unit cell. Secondly,
it is surprising to find that the obtained $T_{N}$ values do not depend on $x$ but only on $\beta$.
However, experimental results have also indicated that $T_N$ does not depend significantly on doping content.
For example, studies on Bi$_2$Fe$_4$O$_9$($\equiv$BiFe$_2$O$_{4.5}$ ; nominal valency of ~Fe$^{3.0+}$), the material recently recognized as a realization of the  Cairo pentagonal lattice, reported a $T_N = 238$~K for single crystals\cite{Ressouche,Gianchinta}, while for polycrystals, $T_N$ was reported to be $\sim260$~K.\cite{Singh,Shamir, Tutov} For mixed valent polycrystalline
BiFe$_2$O$_{4.63}$ with a nominal valency of Fe$^{3.2+}$, which corresponds to a hole doping of $\sim20\%$, 
Retuerto et al. reported a value of $T_N = 250$~K.\cite{Retuerto}
Further, for $x = 1/6$ and $1/3$, since the magnetic structures have no frustration, the zero temperature free energy $f(T=0)=E_{min}$. However, the two magnetic structures have different values of $E_{min}$. From Eq.(\ref{eq:isingunitcell}), we obtain $E_{min}=-(J + \tilde{J})$ for $x = 1/6$, and $E_{min}=-(J + \tilde{J}/2$ for x = 1/3. Thus, even 
with different values of the ground state energies for $x = 1/6$ and $1/3$, our results suggest that $T_N$ depends
only on $\beta$ = $J/\tilde{J}$. 

Finally, the present results also show that the absolute values of the sub-lattice magnetizations at high temperatures($\sim0.25-0.5T_N$/$T_C$ to $T_N$/$T_C$)) depend on the co-ordination of the sites and the value of $\beta$. For example, as can be seen in Fig. 4(a) for $x = 0$ and $\beta=1$, the absolute values of sub-lattice magnetizations of the 3-co-ordinated sites $m_i, i = 1,\cdots,4$ are exactly the same, but for $\sim0.5T_N$ to $T_N$, they are slightly lower than $m_6$, which is a 4-co-ordinated site. Similarly, for the ferrimagnetic phase with $\beta=3$, Fig. 4(b) shows that absolute values of  $m_i, i = 1,\cdots,4$(3-co-ordinated sites) are exactly the same, but for $\sim0.25T_C$ to $T_C$, they are lower than $m_5$ and $m_6$ which are 4-co-ordinated sites. This was also pointed out by Urumov for the ferrimagnetic phase using an exact calculation, although the changes were very small.\cite{Urumov}
For $x=1/6$, the changes are the same as for $x=0$ and $\beta=1$, but the difference between 3 and 4-co-ordinated sites get enhanced on increasing $\beta= 3$. Very interestingly, we see the opposite behaviour for $x =1/3$ compared to $x= 1/6$. For $\beta=1$(Fig. 6(a)), due to selective exchange coupling, the structurally 4-co-ordinated sites become magnetically 2-co-ordinated sites(see Fig. 1(d)), and consequently, the absolute values of $m_5$ and $m_6$ get suppressed compared to $m_1$ and $m_3$(structurally and magnetically 3-cordinated sites) for temperatures between  $\sim0.25T_N$ to $T_N$. In contrast, for $\beta=3$, the difference between absolute values of $m_5$, $m_6$ compared to $m_1$, $m_3$ become smaller as the magnetization profiles get dominated by the larger value of $J$ compared to $\tilde{J}$.

\section{Conclusions}
\label{conclusions}

 In conclusion, we have investigated the spin $S$ = $1/2$ Cairo pentagonal lattice with respect to selective exchange coupling, in a nearest-neighbour antiferromagnetic Ising model. 
We have developed a simple method to quantify geometric frustration in terms of a frustration index $\phi(\beta,T)$, where $\beta$ = $J/\tilde{J}$, the ratio of the two exchange couplings required by the symmetry of the Cairo lattice.
At $T = 0$,  the undoped Cairo pentagonal lattice shows a first order phase transition with antiferromagnetic order  for $\beta \le \beta_{crit}  = 2$, which transforms to a ferrimagnet for $\beta >$ $\beta_{crit}$. 
$\phi(\beta,T = 0)$ exhibits a cusp maximum at $\beta_{crit}$. 
The obtained magnetic structures reveal that the frustration originates in different bonds for the two phases. 
The frustration and ferrimagnetic order get suppressed by selective exchange coupling, and the system shows antiferromagnetic ordering for a virtual doping of $x$ = 1/6 and 1/3. From mean-field calculations, we obtained the temperature-dependent sub-lattice magnetizations for $x$ = $0, 1/6$ and $1/3$. The calculated results were discussed in relation to known experimental results for trivalent Bi$_2$Fe$_4$O$_9$ and
mixed valent BiFe$_2$O$_{4.63}$. The results show the fundamental role of frustration and selective exchange coupling in determining the kind of spin ordering and how they transform in the Cairo pentagonal lattice.

\section{Acknowledgement} We sincerely thank Professor Viktor Urumov, Institute of Physics, Macedonia, for sending us Reference 15.

\end{document}